# Approaches to Modeling the Impact of Cyber Attacks on a Mission

By:

*Alexander Kott, US Army Research Laboratory, USA*

*Mona Lange, University of Lübeck, Germany*

*Jackson Ludwig, The MITRE Corporation, USA*

## Abstract

The success of a business mission is highly dependent on the Communications and Information Systems (CIS) that support the mission. Mission Impact Assessment (MIA) seeks to assist the integration of business or military operations with cyber defense, particularly in bridging the cognitive gap between operational decision-makers and cyber defenders.   Recent years have seen a growing interest in model-driven approaches to MIA. Such approaches involve construction and simulation of models of the mission, systems, and attack scenarios in order to understand an attack's impact, including its nature, dependencies involved, and the extent of consequences.  This paper discusses representative examples of recent research on model-driven approach to MIA, highlights its potential value and cautions about serious remaining challenges.

## Introduction
The success of a business mission is highly dependent on the Communications and Information Systems (CIS) that support the mission. Cyber attacks on CIS degrade or disrupt the performance and completion of the associated mission capability. There is a need for technology and procedures to characterize the impact of a cyber attack on the mission. The term Mission Impact Assessment (MIA) is often used to refer to this characterization problem.

One of the key objectives of MIA is to assist the integration of business or military operations with cyber defense, particularly in bridging the cognitive gap between operational decision-makers and cyber defenders [1].  In other words, MIA supports the ability to determine how an attack on the CIS infrastructure translates into consequences expressed in business, operational terms, and thereby help decision-makers translate operational priorities into cyber defense priorities. [1]  Given a cyber threat and an attack with certain characteristics, cyber MIA is expected to identify the space of impact scenarios: the CIS assets that would be disrupted, the chain of dependencies through which the CIS disruptions will propagate to business functions, and the resulting degradation in quantity and quality of the outputs of the business process affected.

---

[1] This paper is inspired in part by the NATO-sponsored workshop [22] titled "Cyber Attack Detection, Forensics and Attribution for Assessment of Mission Impact," held in Istanbul, Turkey, in July 2015.

MIA can be considered as a sub-field within a much broader and far more mature field of risk management, particularly within the sub-process of risk management called risk identification. [2] This sub-process involves identification of business processes, functions and supporting assets; threats, including strategically thinking and adaptive threats; vulnerabilities that could be exploited by threats; security controls or measures; and – especially relevant to MIA – technical and business consequences or impacts.  In complex, multi-organizational, cyber or cyber-physical systems, such impacts can be multifaceted, distributed in time and space, propagated through poorly understood dependencies, difficult to visualize and anticipate, and even counter-intuitive.  Effective computational approaches to automating or supporting MIA are lacking. This encourages the cyber security community, in particular, to pay increasing attention to cyber MIA as a distinct problem.

Although attracting a distinct and growing body of research (e.g., [3], [4]), cyber MIA remains a nascent field. For example, [5] characterizes mission modeling and mission impact assessment is an emerging field of research, provides a review of the relevant literature, and argues that many current approaches to mission impact assessment typically employ score-based algorithms leading to spurious results.

We observe a growing interest in the MIA research community in a simulation model-driven paradigm [21]. It requires creation and validation of mechanisms of modeling the organization whose mission is subject to assessment, the mission (or missions) itself, and the cyber-vulnerable systems that support the mission. The models are then used to simulate or otherwise portray cyber attacks to understand their impacts.

In the following, we illustrate this trend with two specific examples, and also mention related efforts. The two examples cover a broad range of business domains, and a range of research communities: one relates to the domain of civilian electric power distribution, another – to a military planning enterprise; one is the work of European Union researchers, another – of USA. We point out potential value of such approaches, but also the fact that sufficient evidence of value and feasibility is still lacking; and we sound a cautionary note in discussing a number of significant remaining challenges.

## An Example of a Model: Impact of Cyber Attacks on Power Grids

On an operational level, an electrical grid is a network of power providers and consumers that are connected by transmission and distribution lines with the mission of delivering electricity from suppliers to consumers. For monitoring and control purposes, they are connected to CIS. As recently as the 1990s, many power grid networks were isolated, standalone systems, and the day-to-day functioning of an electrical power grid mainly depended on the correct functioning of physical devices such as transmission and distribution lines, generators, and transformers. However, modern power grids and CIS infrastructures are closely coupled. Previously isolated power grids are increasingly integrated with CIS at power utilities, including public infrastructures, in order to increase business efficiency, effectiveness, and reduce operational costs. This has led to modern power grids becoming large networks consisting of thousands of network devices and applications. The operability, performance, or reliability of an application may depend on multiple network services spanning multiple network devices and sub-networks of an infrastructure.

Both the U.S. Department of Homeland Security (DHS) and the U.S. Department of Energy reported an increase in frequency and sophistication of cyber attacks on electricity systems [6]. A growing number of host and network intrusion detection systems and firewalls are deployed in electricity systems, leading to a high number of detected low-level events. Managing these low-level events and assessing their potential operational impact is critically important [7]. A deeper understanding of potential impacts resulting from a successful cyber attack is required, especially for the development of trustworthy smart grids.

This was one of the goal of the PANOPTESEC project (2013-2016) funded by the Seventh Framework Program for Research (FP7) of the European Commission. The resulting PANOPTESEC prototype demonstrated a continuous monitoring and response capability to detect, prevent, manage and react to cyber incidents in real-time. For the purposes of this paper, given a suspected attack, PANOPTESEC evaluates their operational impact, i.e., performs MIA, among other capabilities [8].

In fact, PANOPTESEC is one of several recent, related projects. For example, The CRISALIS project (Critical Infrastructure Security Analysis, http://www.crisalis-project.eu/), was funded by in 2012-2015. It focused on three themes: (i) securing the systems, (ii) detecting the intrusions, (iii) the "post-mortem" analysis of successful intrusions. The bulk of the results produced under CRISALIS relate to intrusion detection in SCADA and industrial control systems, as opposed to mission impact assessment. Although CRISALIS research did not extend into characterizing explicit impacts on business functions and processes, it did touch on interests of MIA where it explored approaches to detecting attacks against ICS devices by observing changes in the industrial process variables (e.g., [9]).

Also funded by the FP7, for years 2014-2016, the HyRiM (Hybrid Risk Management for Utility Networks, https://hyrim.net/) project aimed to identify and evaluate "Hybrid Risk Metrics" for assessing and categorizing security risks in interconnected networks: the utility network physical infrastructure, consisting of, e.g. gas, water pipes or power lines, and the utility's control network including SCADA (Supervisory Control and Data Acquisition) networks and business and information systems. The project provide risk assessment tools that is based on a sound and well-understood mathematical foundation. The bulk of the HyRiM research has concentrated on formal mathematical, including game-theoretic, models of risks.

Perhaps the most direct ancestor of PANOPTESEC was the VIKING project (Vital Infrastructure, Networks, Information and Control Systems Management, http://cordis.europa.eu/project/rcn/88625_en.html) financed by EU in 2008-2011. Its key objective was to investigate the vulnerability of SCADA systems and the cost of cyber attacks on society, focusing on systems for transmission and distribution of electric power. It took a model-based approach to investigating SCADA system vulnerability. Models were defined for the SCADA system, for the electrical process as well as of for the society that is dependent on the electricity supply. The models were linked to assess the propagation of consequences from a cyber attack all the way to the impact – expressed as monetary loss -- for the society. The results laid a foundation for further exploration in the PANOPTESEC project.

As a case study, the PANOPTESEC consortium set up a testbed, which is an authentic replication of Italian water and energy distribution company's corporate enterprise systems and SCADA system. This allows for testing PANOPTESEC in an operational environment, and for experimenting with cyber attackers who are able to penetrate computer systems, tamper with the accuracy of information, and shut down network services.

Figure 1 illustrates how an enterprise network and an operational network are linked in power grids. An enterprise network consisting of a primary control center is linked to an operational network consisting of two substations and an advanced metering infrastructure, represented by a smart home with a smart meter, heating, and a thermostat. The linkage of enterprise and operational networks is due to sensor measurements and control commands from power system operators in control rooms within an enterprise network being relayed over communication networks from or to a power grid's operation network. Clearly, there are multiple, complex dependencies between business functions and tasks, devices and applications, and network services.

An important finding of this project was the recognition that manual modeling of dependencies is prohibitively expensive in complex enterprises where responsibilities and knowledge are scattered across multiple departments or even third parties. Thus, a key output of our project was the development of an automated approach to automatically learning network dependencies based on network traffic, and then deducing higher-level information about a network's mission based on network services and applications (Figure 2).

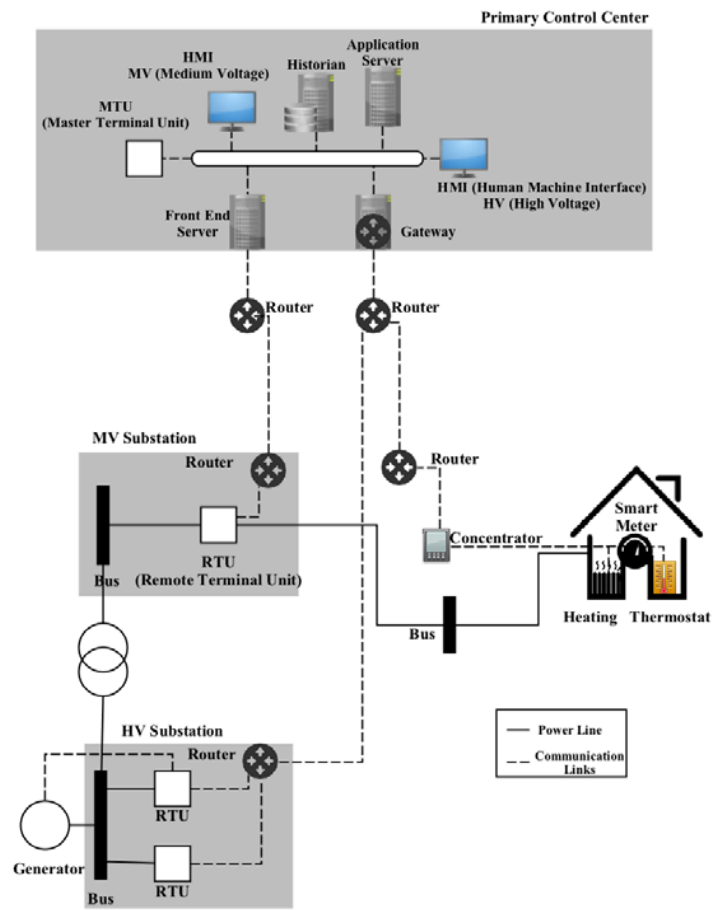

*Figure 1. Even the simplest schematics highlights the diversity and complexity of cyber and physical elements of a modern electric grid.*

Automated model development relies in part on network services dependency discovery, for which a number of approaches are known, e.g., [10]. Discovery of indirect dependencies between services is particularly challenging. In our PANOPTESEC project we analyze communication patterns and derive indirect dependencies based on "similar" temporal patterns of communications between two given pairs of services. Normalized cross-correlation is used heuristically to quantify the degree of similarity (details are in [8]). This heuristic technique has been shown experimentally [11] to outperform several alternative approaches such as [10] in terms of recall and precision of indirect dependencies discovered.

The dependencies identified via the automated method are used in part to construct a mission model, e.g., Figure 2. Based on the mission model we are able to detect what applications are potentially impacted by an attack. Assuming the operational network is sufficiently modeled in the infrastructure model, mission impact assessment includes estimating the potential loss of electric power and whether it could lead to a black out or brown out in the monitored infrastructure.

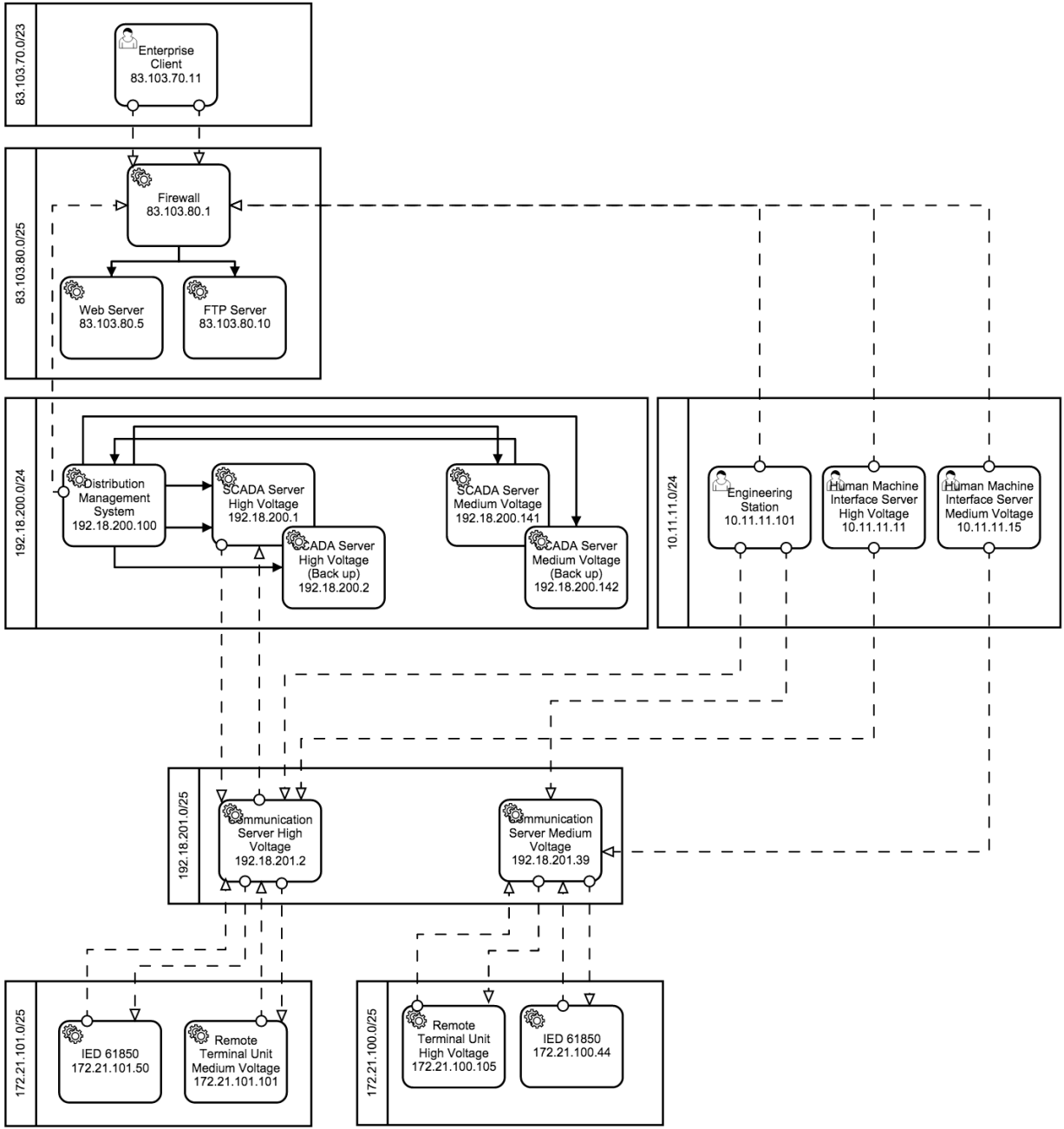

*Figure 2 An example of a high level view of an automatically derived mission models. Swim lanes represent sub networks, network devices are represented by tasks and a human silhouette marks client network devices.*

By comparing automatically derived mission models with results based on human input, we found automatically derived mission models to provide a more detailed understanding of workflows in the network, and also to discover a surprisingly high number of hidden network dependencies that heretofore were not identified by human operators. Unsurprisingly, network administrators found these automatically discovered, previously unknown network dependencies of great interest.

As an example, consider automatically generated diagram of significant relations between a subset of services within the PANOPTESEC testbed (Figure 3). When PANOPTESEC researchers asked the operators, they explained that HMIs such as msoz22, msoz17 and msoz19 are used to communicate with the medium voltage substations (including mferp2, and various TTYs connected to mferp2) through the communication server muel2.

However, our automated analysis has shown a peculiar fact that was unknown to the operators: all HMIs (for example, msoz22, msoz17 and msoz19) were configured to contact muel1 first, and then contact muel2 if muel1 is unavailable. At the same time, muel1 was configured to be the backup of muel2. Therefore, muel1 normally rejects requests from HMIs, because muel1 can check and determine that muel2 is available. Then, after the rejection by muel1, the HMIs send requests to muel2.

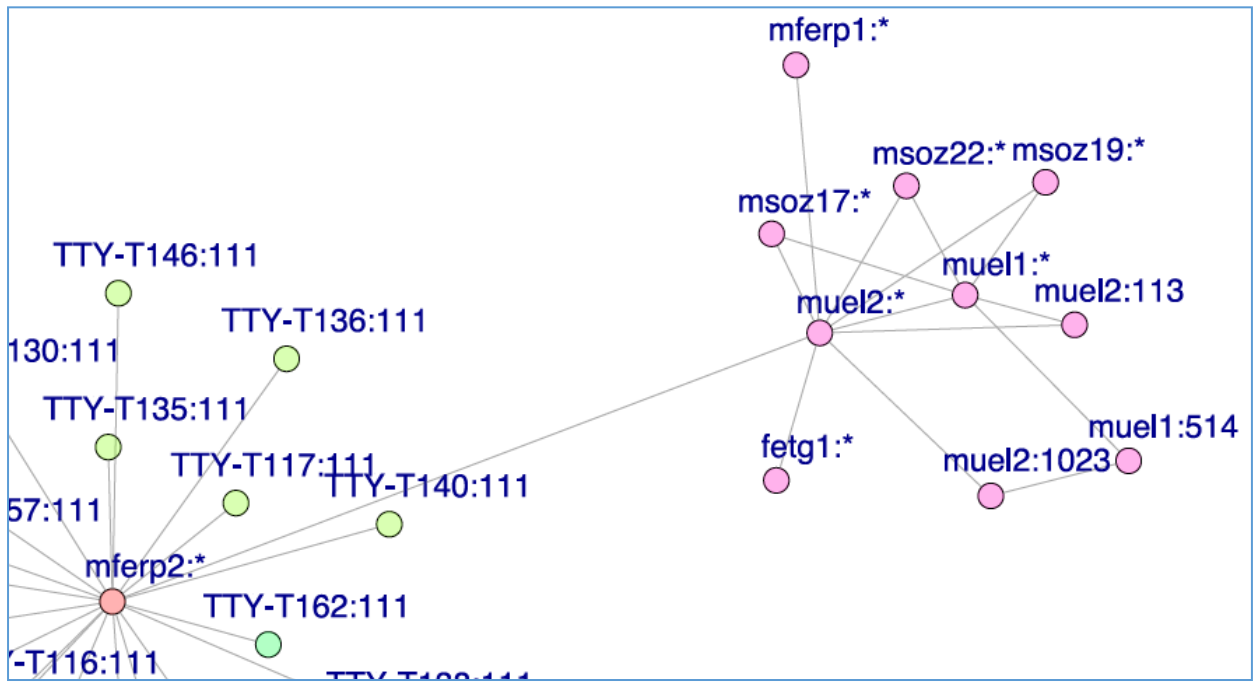

Figure 3These dependencies between services were identified automatically, and revealed unknown dependencies due to mis-configurations.

In effect, automated analysis discovered that there was unexpected, unnecessary-- and potentially exploitable for malicious purposes – heavy volume of communications between HMIs (msoz22, msoz17 and msoz19) and muel1, even though all the communications between HMIs and mferp2 eventually occurred through muel2. After automated analysis discovered this fact, the network operators adjusted the configurations accordingly. A prototype tool for dependency analysis has been developed as a part of this research project and is being released under an open source license [8].

To be sure, not everyone agrees that automated dependency analysis is adequate for constructing a comprehensive mission model for MIA. For example, a group of researchers mentioned in the next section of this paper specifically evaluated several tools for automated dependency discovery and found them inadequate for the task; they proceeded to build the models manually, relying on the input of Subject Matter Experts (SMEs). Furthermore, it should be noted that the PANOPTESEC models has never been fully validated, and the business value of the modeling and simulation with such model is yet to be rigorously confirmed.

# Another Example of a Model: Impact of Cyber Attacks on a Military Air Operations Center

In 2015, the U.S. Department of Defense funded a team of research organizations to develop a research prototype. The goal was to explore the feasibility of modeling and simulating concurrently the operational and cyber domains, and translate the impact of cyber events into quantifiable impacts to the execution of an operational mission.  The outcome of this research effort was a prototype called Analyzing Mission Impacts of Cyber Actions (AMICA).

Understanding mission impact due to cyber attacks requires bringing together layers of information from numerous sources.  At the lower layers, network topology, firewall policies, intrusion detection systems, system configurations, vulnerabilities, etc., all play a part.  Similarly, network devices and applications also need to be mapped to mission requirements.  Because missions are highly dynamic, key network devices and applications likewise become dynamic.  To address this, time-dependent models of mission flow and cyber actions (attack and defense) are necessary.  AMICA supports exploration and experimentation of the mission impacts of cyber attacks through a flexible, extensible, modular, multi-layer modeling system for quantitative assessment of operational impacts of cyber attacks on mission performance [12].

As a case study, AMICA was used to examine potential cyber impacts to an Air & Space Operations Center (AOC).  At the risk of oversimplification, an AOC is responsible for developing the daily mission priorities and flight schedules for all the aircraft involved in a military campaign [13].  To conduct its mission, an AOC requires a large team of people working around the clock performing a planning and decision-making process, with a deep reliance on CIS.

AMICA consists of four main components: a mission process model, a cyber adversary process model, a cyber defender process model, and CIS infrastructure model (Figure 3).  The mission, cyber adversary, and cyber defender are all modeled in terms of their respective business (operational) processes using the Business Process Model and Notation (BPMN) standard [14].  The CIS infrastructure is modeled through the use of both directed graphs and topological graphs.  Behavioral and temporal aspects of the AOC processes (workload, workflow, timing constraints, required resources, decisions, etc.) are implemented through executable process models and stochastic discrete-event simulation.  Structural and functional aspects related to the AOC infrastructure (environmental constraints, mission and system dependencies, vulnerabilities, etc.) are maintained through databases of graph models.  Each component of AMICA is decoupled from the rest to provide modularity and independence, and interacts via shared interfaces with the CIS infrastructure model.  This allows inputs at both the operational and cyber layers to influence the behavior of the CIS layer and produce a combined effect on mission performance.

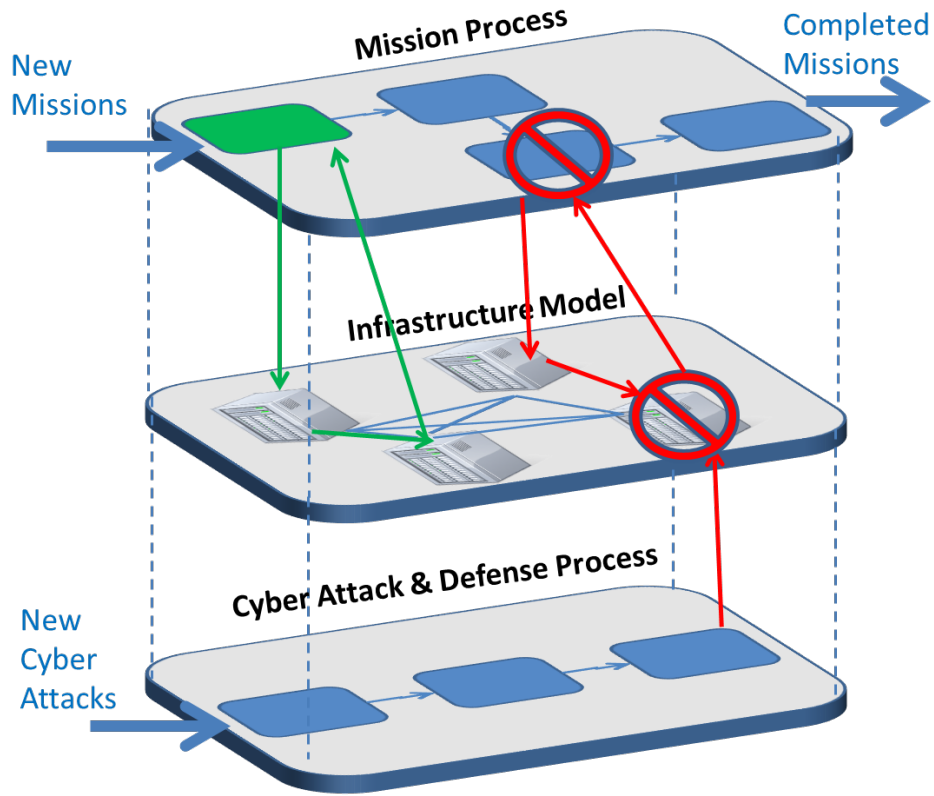

*Figure 4. The overall model includes a mission process model, a cyber adversary process model, a cyber defender process model, and a CIS infrastructure model.*

AMICA models the progress of each aircraft flight through the AOC's planning process, dependent on the state of the cyber infrastructure.  Cyber attacks themselves are modeled as stochastic steps within the attacker's workflow (modeled using the cyber kill chain [15]) and follow a pattern of gaining access to the network, lateral movement, and exploitation of the target device.  Gaining access is done via spearphishing, where end-user nodes have a probability of falling victim to a spearphishing attempt.  Lateral movement through the network is based on scanning the network for vulnerable devices that the adversary has the capability to exploit.  Once the target device has been reached, the attacker creates a confidentiality, integrity, or availability impact on that system.  Availability attacks slow down or stop execution of impacted mission activities.  Confidentiality and integrity attacks do not slow execution, but may cause the AOC personnel to perform rework (if they are aware of the attack), or allow corrupted data to appear on flight plans.  The duration of the attack depends upon how quickly the defender detects the presence of an attack, performs forensics, finds all the machines that have been compromised, and completes the remediation process.  The consequence of these cyber attacks are reported in terms of mission-level metrics.  In the case of the AOC, the number of mission plans developed and the number of flights flown are typical metrics of interest.

Using simulation to explore the impacts over time of cyber attacks on a specific AOC mission scenario revealed impacts that ranged from: no impact at all (because of low workload intensity and redundant systems), to mission plans being delayed by days (because personnel needed to re-validate data after discovering a breach), to unknowingly having an entire day's worth of flights modified by an adversary (because an attack took place after the last consistency checks were made).

We observed that randomly timed attacks against critical nodes within the CIS infrastructure, while disruptive, were not devastating to the mission. On the other hand, attacks against key process steps conducted by attacking the same portion of CIS at the right moment during the process caused severe mission impacts. We conclude that cyber defenders need to assume that an advanced cyber adversary will likely target process vulnerabilities using cyber vulnerabilities as a vector.

Several findings of this modeling effort were surprising, and would not be possible to obtain without a comprehensive, system-wide modeling approach. In one example, two simulated attacks were run against systems supporting an important planning process and the duration of each was varied (from hours to weeks) to determine the maximum tolerable attack duration. In one attack (Figure 5), system performance was degraded by 50%. This showed no impact since the users were able to keep pace with their workload even when using slow systems. In the second attack, the external network connection was disrupted, cutting off access to teammembers located at remote locations. In this case, the local users were nearly able to keep up with the workload, and only for outages exceeding three weeks did the reduction in completed flight plans reach a significant level (defined as a 10% or greater reduction). Since outages of such a duration are unlikely, this combination of process, workforce, and systems can be viewed as tolerant of attacks.

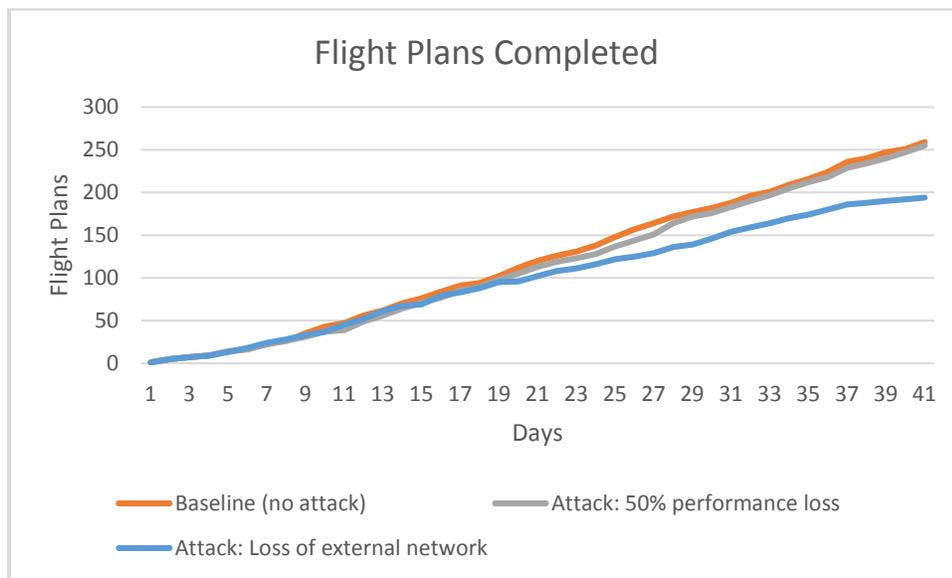

*Figure 5. The model shows unexpectedly limited impact of cyber attacks if the workload of the planners is sufficiently low.*

In another example, attacks resulting in data modification were targeted at varying times against systems supporting a critical flight planning process. Static dependency analysis would correctly show that attacks against these systems would cause a severe impact to operations. However, simulation results showed that only attacks that took place after the last consistency checks (approximately halfway into the day) were made had an impact to the mission, providing a discrete window vulnerability for defenders to focus on each day. The impact of attacks – expressed as a fraction of plans with undetected malicious modifications – ranged from the minimum of 3% to 100% on the agility and proficiency of the attacker modeled.

We also found that such comprehensive modeling does not come cheaply. The effort involved in applying these approaches with current techniques to a new mission area requires months of effort by knowledgeable SMEs. Once modeled, scenarios and simulation runs can be done quickly but, a rigorous analysis of the data may take a few weeks. Attack trees, dependency graphs, and vulnerability scan data can each provide partial solutions identifying where or how a cyberattack might take place. But none can quantify or place bounds the mission impact since those approaches do not include time. Governments and large enterprises facing substantial cyber threats may benefit from the additional detail that simulation can provide; however, this is not an approach that is well suited to casual users given the present state of technology.

Our current prototype took 3 person-years to develop and contains only the features and datasets needed to support demonstration of the approach. A production-quality system would require further significant enhancements to data import, execution speed, scenario design, and user interfaces to make the software suitable for non-developers. The majority of the recurring effort in applying this approach is in collecting the necessary data about systems and processes, and the pertinent dependencies. As another example of a recent experience, developing cyber dependency graphs for a large military organization with users, missions, systems, and networks spanning multiple locations, took a team of developers over 7 person-years to complete, with further ongoing effort required to maintain the dependency graphs as changes are made to the infrastructure over time. A manual, SME-based mapping approach had to be used because an AMICA-related development team found that automated tools were insufficiently reliable in finding true dependency relationships – contrary to PANOPTESEC's experience. The automated tools showed poor recall and precision, i.e., too many missed dependencies and too many false dependencies.

## Mission Impact Assessment Problem Formulation

Having considered two illustrative examples, we now discuss major remaining challenging.

To begin, appropriate formulation of a problem is the key to its successful solution. What constitutes a successful solution depends in turn on who are the users of the solution. For example, decision-makers need decision support at an appropriately abstracted level; they are much less interested in technical details. For these reasons, future techniques of MIA may specifically focus on supporting the cyber security decision making process, and particularly on tools that teach, train, and support decision makers.

Determining the correct users, however, depends on knowing where MIA belongs in the broader scheme of things. One way is to consider MIA as a part of the big control loop that strives to keep the controlled "plant" – the mission – within the prescribed space of secure states [16]. The output of this controller is a set of corrective actions designed to keep the plant in the secure state.

In this formulation, MIA is the component of the control system that measures how much the plant has deviated or will deviate from the desired state. Once we say "how much," a formal quantification of a utility function is needed. Some of current approaches to MIA are based on heuristic scoring. To oversimplify, the assessor sums up the "impact points" and declares that the total impact on the mission is, let's say, 73 points. Similarly, to say that "The mission impact is 70% failure" is very difficult to interpret. E.g., even with 70% of mission failing (whatever that means), the operator may still be able to reach a key goal.

A way to express quantitative output of MIA would be to measure mission impact as a reduction in tangible attributes of the system, such as the network bandwidth, delay, or power use. Yet another approach would be to quantify the distance from the achievable states to desired states, e.g., via the cost of the corrective actions that would bring the plant to the desired, secure state. All this suggests that a formal language, a formal mathematics of mission security would be highly desirable to give MIA a solid quantitative foundation.

Appropriate formulation of the MIA problem also requires choosing the right level of abstraction. When formulated and solved at a very abstract level, the solution may not give adequate insights into what actions – often very specific and detailed – need to be considered. On the other hand, when formulated at a very detailed level, the problem demands a very intricate model that is far too expensive to construct. Arguably, a shift must occur from the enterprise scale problem to more meaningful tactical scale. One argument for more detailed formulations is that seemingly small attacks on mission activities can have large effects, as confirmed by simulation studies [12].

In addition to the control-theoretic style of the MIA problem formulation, we should not overlook the game theoretic (or game-playing simulation) perspective. A related and appropriate style of problem formulation could be the robust control with adversarial inputs. Because full information is not normally available, the problem should be formulated as a partial information game.

## Model Content

The fundamental components of a model required for MIA include the models of the Organization, its Business Processes (often decomposed into functions and tasks), the Missions executed through the Business Processes, and the CIS that support the missions. Relations, influences and dependencies – quantitatively characterized – between all these entities and their sub-entities need to be modeled. Even the physical environment of a mission may need to be modeled, as well as the sensors and actuators that sense and affect the environment, because they also can be subjects of a cyber or deception attacks.

Because the MIA problem is so fundamentally adversarial in nature, it was widely recognized that one needs a comprehensive model for Adversary characterization and behavior understanding and prediction; the model should also include Environment Property, Attacks Property and Target Property including modeling of these three elements and relations and interactions between them.

In addition to describing the structure of the problem, models must capture its dynamics. There are several very different meanings of dynamics in MIA models. First, the structure itself changes rapidly. For example, the servers supporting a mission might be taken down for maintenance and then brought up on line again, or reassigned to another mission. The model would need to be updated continually to reflect such changes. Second, when a cyber attack impacts a mission, the defenders and operators of the mission and supporting systems often show remarkable ability to work around the established process, i.e., to redesign the business process rapidly and radically. Third, even in a very static structure of the business, actions are dynamic – they start, proceed and stop in time. This dynamics also has to be captured in a model. Fourth, the characteristics of components and relations within the model may change depending on the context. For example, criticality of systems change during different missions sharing same systems.

## Models of Adversary

Mission impact has to be considered in the context of what impact the adversary desires. If we know, or are able to estimate, the intents, motivations and anticipations of the adversary, the impact of the adversary on our missions, or the intended impact, would be easier to assess. E.g., in the AMICA system, the model of the attacker (agility and skills level) was shown to influence strongly the mission impact. It should be noted that here we consider the adversary rather abstractly; in particular, we do not assume that the adversary can be modeled as an individual human or a collection of individual humans. That perspective will be discussed a bit later.

A model for adversary characterization, behavior understanding, and prediction should be sufficiently comprehensive. In particular, the model should include properties of the environment in which the cyber conflict occurs; the properties of the attacks and targets that are available to the adversary; and relations and interactions between all such elements; all this in addition to the properties of the adversary itself. Naturally, properties and characteristics of the adversary are often unknown or uncertain. Modeling tools should allow representing such uncertainties.

A powerful determinant of adversary behavior is his expectations of our response. Thus, it is important to understand the role of deterrence – the measures we can take to prevent hostile actions by adversary – in a cyber conflict. An adversary model should help answer questions like: what does the adversary want to do, and what they expect us to do?

## Models of Humans

When the adversary is an individual human, or a group of individuals that we find appropriate to model individually, we should consider techniques of cognitive modeling of individual human minds. Such models can help predict how the cyber attacker formulates his/her goals, and thereby tell us about the intended or actual mission impact of the adversary actions. Cognitive modeling tools like ACT-R are beginning to be used for modeling behaviors of cyber attackers [17]. Still, this research area tends to be rather immature.

Models of defenders should not be overlooked either. To assess the likely mission impact, we need to know how a human cyber defender reacts to cyber attacks. Errors committed by defenders determine the extent of mission impact. A defender may fail to recognize a threat and to take appropriate actions (or take the wrong action based on imperfect information), thereby enabling a greater mission impact. A defender may fall a victim to deception committed by an attacker [18]. A defender may fail to undertake a suitable work-around when a mission is impacted. A defender may also misinterpret the mission impact when it occurs. All this is highly relevant to the MIA problem.

Whether one models an attacker or a defender, the model needs to be rich enough to reflect "irrational" aspects of human cognition, such as cognitive biases. These aspects are particularly important in a high-pressure, high-tempo, non-intuitive world of cyber operations. Impact of dynamic learning must be considered to account for rapid evolution of knowledge in cyber conflicts. Game-theoretic approaches should be included in order to account for the highly adversarial nature of cyber operations. Because both the attacker and the defender operate often with very limited awareness of each other actions, situational awareness of both should be modeled. The importance of situational awareness in achieving impact on the opponent's mission cannot be overlooked.

In many cases, however, both the defender and the attacker are best modeled not as individual human cognitive actors, but rather as organizations. Organizational modeling is studied by a community of researchers in the political science field that is distinct from the community of cognitive modelers. It would be worth exploring how that community might help solving the MIA problem.

## Model Construction

The current practice of constructing models for MIA is almost entirely manual in nature. As such, model construction is very time consuming, expensive, difficult to document, to inspect and to validate. Maintenance of such models – also manual – is also expensive. Quantitative characterization of dependencies between, for example, business functions and supporting technical assets, is largely a matter of asking the presumed Subject Matter Experts for a number, such as a conditional probability. The guessing of such numbers of SME is expensive and the verity of numbers is doubtful.

Still, manual construction of models for MIA problems appears feasible, even if expensive. For example, both PANOPTESEC and AMICA are comprehensive MIA modeling and simulation systems with a relatively fully implemented business model. To a degree, both rely on manually crafted models (AMICA more so).

Some tools exist that allow essentially manual yet computer-aided construction of business models. Widely available Business Process Management (BPM) tools fall into this category. Ideally, however, we would like to see the bulk of MIA models constructed automatically, perhaps by observing a business process and its cyber defense operations, and automatically learning or inferring a model. An approach has been described where a significant part of a model was automatically derived from the observed network flows.

## Conclusions

Business effectiveness of cyber defense depends to a large degree on the ability of the business to assess – systematically and quantitatively – the impact of cyber attacks on the business mission. This problem of Mission Impact Assessment will not be solved by an ad-hoc muddle of compliance check lists, forensic investigations and expert opinions. Like in most mature technical and management fields, the problem will require comprehensive models [23]. To be sure, challenges of building such a model are formidable. They range from a formulating the model around the right decision-making needs and at the right level of detail; finding effective means of representing complex adversarial and human-cognition aspects of the domain; and developing cost-effective approaches to constructing and validating a model. While experience with recent research projects suggests feasibility and potential utility of developing such models, decisive evidence of their benefits awaits further research and practical deployments.

# References

[1] A. Kott, Cyber Defense and Situational Awareness, C. Wang and R. Erbacher., Eds., New York: Springer, 2014.

[2] ISO/IEC, Information technology -- Security techniques-Information security risk management ISO/IEC FIDIS 27005, ISO/IEC, 2008.

[3] G. Jakobson, "Mission cyber security situation assessment using impact dependency graphs," in *Proceedings of the 14th International Conference on Information Fusion*, 2011.

[4] S. Musman, A. Temin, M. Tanner, D. Fox and B. Pridemore, "Evaluating the impact of cyber attacks on missions," in *Proceedings of the International Conference on Information Warfare and Security*, 2010.

[5] A. Motzek and R. Möller, "Context- and bias-free probabilistic mission," *Computers & Security,* vol. 65, pp. 166-186, 2017.

[6] United States Department of Energy, "Quadrennial Technology Review," 2015.

[7] S. Sridhar, A. Hahn and M. Govindarasu, "Cyber–physical system security for the electric power grid," *Proceedings of the IEEE,* vol. 100, no. 1, pp. 210-224, 2012.

[8] M. Lange, R. Moeller, G. Lang and F. Kuhr, "Event Prioritization and Correlation Based on Pattern Mining Techniques," in *Proceedings of the 14th International Conference on Machine Learning and Applications*, Miami, FL, USA, 2015.

[9] A. Carcano, A. Coletta, Guglielmi, M. Masera, I. Fovino and A. Trombetta, "A multidimensional critical state analysis for detecting intrusions in SCADA systems," *IEEE Transactions on Industrial Informatics,* vol. 7, no. 2, pp. 179-186, 2011.

[10] X. Chen, M. Zhang, Z. M. Mao and P. Bahl, "Automating network application dependency discovery: Experiences, limitations, and new solutions," in *Proceedings of the USENIX Symposium on Operating Systems Design and Implementation*, 2008.

[11] M. Lange, F. Kuhr and R. Möller, "Using a Deep Understanding of Network Activities for Network Vulnerability Assessment," in *Proceedings of the 1st International Workshop on AI for Privacy and Security*, 2016.

[12] S. Noel, J. Ludwig, P. Jain, D. Johnson, R. Thomas, J. McFarland, B. King, S. Webster and B. Tello, "Analyzing Mission Impacts of Cyber Actions," in *NATO IST-128 Workshop on Cyber Attack Detection, Forensics and Attribution for Assessment of Mission Impact*, Istanbul, 2015.

[13] R. Thompson, "Realizing Operational Planning and Assessment in the Twenty-First-Century Operations Center," *Air & Space Power Journal,* vol. 27, no. 2, p. 107, 2013.

[14] Object Management Group, "Business Process Model and Notation," [Online]. Available: http://www.bpmn.org/.

[15] E. Hutchins, M. Cloppert and R. Amin, "Intelligence-Driven Computer Network Defense Informed by Analysis of Adversary Campaigns and Kill Chains," Lockheed Martin, 2010. [Online]. Available: http://www.lockheedmartin.com/content/dam/lockheed/data/corporate/documents/LM-White-Paper-Intel-Driven-Defense.pdf.

[16] A. Kott, "Towards fundamental science of cyber security," in *Network Science and Cybersecurity*, New York, Springer, 2014, pp. 1-13.

[17] C. Gonzalez, N. Ben-Asher, A. Oltramari and C. Lebiere, "Cognition and technology," in *Cyber Defense and Situational Awareness*, Springer International Publishing, 2014, pp. 93-117.

[18] A. Kott, Information warfare and organizational decision-making, Artech House, 2006.

[19] M. Lange, R. Moeller, G. Lang and F. Kuhr, "Event Prioritization and Correlation based on Pattern Mining Techniques," in *14th International Conference on Machine Learning and Applications*, Miami, 2015.

[20] Internet Assigned Numbers Authority, "Service Name and Transport Protocol Port Registry," 2 December 2015. [Online]. Available: http://www.iana.org/assignments/service-names-port-numbers/service-names-port-numbers.xhtml.

[21] Lange, Mona, Alexander Kott, Noam Ben-Asher, Wim Mees, Nazife Baykal, Cristian-Mihai Vidu, Matteo Merialdo, Marek Malowidzki, and Bhopinder Madahar. "Recommendations for Model-Driven Paradigms for Integrated Approaches to Cyber Defense." arXiv preprint arXiv:1703.03306 (2017).

[22] Kott, Alexander, Nikolai Stoianov, Nazife Baykal, Alfred Moller, Reginald Sawilla, Pram Jain, Mona Lange, and Cristian Vidu. "Assessing Mission Impact of Cyberattacks: Report of the NATO IST-128 Workshop." arXiv preprint arXiv:1601.00912 (2016).

[23] Kott, Alexander. "Towards fundamental science of cyber security." In Network Science and Cybersecurity, pp. 1-13. Springer New York, 2014.